\documentclass[12pt]{article}
\usepackage{epsfig,cite,amsmath,amssymb,comment,a4p}
\usepackage{color}

\begin{document}

\textheight=23.cm
\bibliographystyle{unsrt}      
\date{}

\def\jour#1#2#3#4{{#1} {\bf#2}, #4 (19#3)}
\def\jourm#1#2#3#4{{#1} {\bf#2} (20#3) #4}
\def\appj#1{{#1}}

\def\EPJ{Eur. Phys. J. {C}}

\def\PRp{Phys. Reports}
\def\PRD{Phys. Rev. {D}}
\def\PRC{Phys. Rev. {C}}
\def\IJ{Int. J. Mod. Phys. {A}}
\def\ML{Mod. Phys. Lett. {A}}
\def\JP{J. Phys. {G}}
\def\AP{Acta Phys. Pol. {B}}
\def\NIM{Nucl. Instr. Meth. {A}}
\def\CP{Comp. Phys. Comm.}

\def\etal{{\em et al.\/}}

\def\tmw4j{$\mw4j$}
\def\delq{\delta Q}
\def\nwp{\newpage}
\def\bi{\bibitem}
\def\vs{\vspace*}
\def\hs{\hspace*}
\def\ct{\cite}
\def\be{\begin{equation}}
\def\ee{\end{equation}}
\def\bea{\begin{eqnarray}}
\def\eea{\end{eqnarray}}
\def\la{\label}
\def\bc{\begin{center}}
\def\ec{\end{center}}
\def\al{\langle}
\def\ar{\rangle}
\def\leq{\leqslant}
\def\geq{\geqslant}
\def\lssim{\stackrel{<}{_\sim}}
\def\gtsim{\stackrel{>}{_\sim}}

\def\ea{{\sl et al.}}
\def\eg{{\sl e.g.}}
\def\et{{\sl etc.}}
\def\ie{{\sl i.e.}}
\def\va{{\sl via }}
\def\vrs{{\sl vs. }}

\def\ep{e$^+$e$^-$ }
\def\z{Z$^0$}
\def\pT{p_T}
\def\phi{\Phi}
\def\yf{y$$\times$$\phi}
\def\yp{y$$\times$$\pT}
\def\fp{\phi$$\times$$\pT}
\def\3d{y$$\times$$\phi$$\times$$\pT}
\def\od{single-particle }

%
%
\def\err{uncertainties }
\def\errp{uncertainties}
\def\intp{intermittency}
\def\BE{\mbox{\sc BE}}
\def\mom{moments }
\def\momp{moments}
\def\cum{cumulants }
\def\cump{cumulants}
\def\mupa{multiparticle }
\def\flus{fluctuations }
\def\cors{correlations }
\def\flup{fluctuations}
\def\corp{correlations}
\def\gen{genuine }
\def\phs{phase space}
\def\psh{phase-space}
\def\MC{Monte Carlo }
\def\HW{{\sc Herwig} }
\def\JT{{\sc Jetset} }
\def\OP{OPAL }
\def\DE{DELPHI }
\def\col{Collaboration}

\def\PYTHIA{{\sc Pythia}}

\def\fig{Fig. }
\def\fgs{Figs. }
\def\rfl{Ref. }
\def\rfs{Refs. }
\def\frm{Eq. }
\def\fre{Eqs. }
\def\aver#1{\langle#1\rangle}
\newcommand{\bfp}{\mathbf{p}}
\newcommand{\bfr}{\mathbf{r}}

\newcommand{\ZF}[3]{Z. Phys. {\bf C{#1}} ({#2}) {#3}}
\newcommand{\ZP}[3]{Z. Phys. {\bf C{#1}} ({#2}) {#3}}
\newcommand{\PL}[3]{Phys. Lett. {\bf {#1}} ({#2}) {#3}}
\newcommand{\MPL}[3]{Mod. Phys. Lett. {\bf {#1}} ({#2}) {#3}}
\newcommand{\PR}[3]{Phys. Rep. {\bf {#1}} ({#2}) {#3}}
\newcommand{\NP}[3]{Nucl. Phys. {\bf {#1}} ({#2}) {#3}}
\newcommand{\PRL}[3]{Phys. Rev. Lett. {\bf {#1}} ({#2}) {#3}}
\newcommand{\PRV}[3]{Phys. Rev. {\bf {#1}} ({#2}) {#3}}
\newcommand{\SJNP}[3]{Sov. J. of Nucl. Phys. {\bf {#1}} ({#2}) {#3}}
\newcommand{\APP}[3]{Acta Phys. Pol. {\bf {#1}} ({#2}) {#3}}

\renewcommand{\ZF}[3]{Z. Phys. {\bf C{#1}} ({#2}) {#3}}
\renewcommand{\ZP}[3]{Z. Phys. {\bf C{#1}} ({#2}) {#3}}
\renewcommand{\PL}[3]{Phys. Lett. {\bf {#1}} ({#2}) {#3}}
\renewcommand{\MPL}[3]{Mod. Phys. Lett. {\bf {#1}} ({#2}) {#3}}
\renewcommand{\PR}[3]{Phys. Rep. {\bf {#1}} ({#2}) {#3}}
\renewcommand{\NP}[3]{Nucl. Phys. {\bf {#1}} ({#2}) {#3}}
\renewcommand{\PRL}[3]{Phys. Rev. Lett. {\bf {#1}} ({#2}) {#3}}
\renewcommand{\PRV}[3]{Phys. Rev. {\bf {#1}} ({#2}) {#3}}
\renewcommand{\SJNP}[3]{Sov. J. of Nucl. Phys. {\bf {#1}} ({#2}) {#3}}
\renewcommand{\APP}[3]{Acta Phys. Pol. {\bf {#1}} ({#2}) {#3}}
%
%
\newcommand{\WSCP}{World Scientific, Singapore}

\newcommand{\MPERICE}{\rm Proc. Europ. Study Conf. on Partons and Soft Hadronic
 Interactions, Erice, ed.~R.T.~Van~de~Walle (\WSCP, 1982)}

\newcommand{\FESTHOVE}{Festschrift L.~Van Hove,
eds.~A.~Giovannini and W. Kittel (\WSCP, 1990)}
\newcommand{\MPGOA}{\rm Proc. Xth Int. Symp. on Multiparticle Dynamics, \
Goa 1979, eds.~S.N. Ganguli, P.K. Malhotra and A. Subramanian (Tata Inst.)}

\newcommand{\MPLUND}{\rm Proc. XV Int. Symp. on Multiparticle Dynamics,
Lund, Sweden 1984, eds.~G.~Gustafson and C.~Peterson  (\WSCP, 1984)}
%
\newcommand{\MPISRAEL}{\rm Proc. XVI Int. Symp. on Multiparticle Dynamics,
Jerusalem, Israel, Sweden 1985, ed.~J.Grunhaus
(Editions Fronti\`eres, France and \WSCP, 1985)}

\newcommand{\MPTASHKENT}{%
\rm Proc. XVIII Int. Symp. on Multiparticle Dynamics, Tashkent, USSR,  1987,
eds.~I.~Dremin and K.~Gulamov (\WSCP, 1988)}

\newcommand{\MPARLES}{%
\rm Proc. 19th Int. Symp. on Multiparticle Dynamics, Arles, 1988,
eds.~D.~Schiff and J.~Tran~Thanh~Van (Editions Fronti\`eres, France
and \WSCP, 1988)}

\newcommand{\MPHOLMECKE}{%
\rm Proc. XX Int. Symp. on Multiparticle Dynamics, Gut~Holmecke, Germany, 1990,
eds.~R.~Baier and D.~Wegener  (\WSCP, 1991)}
%
\newcommand{\MPWUHAN}{\rm Proc. XXI Int. Symp. on Multiparticle Dynamics, 
Wuhan, China, 1991, eds.~Y.F. Wu and L.S. Liu (\WSCP, 1992)}
%
%
\newcommand{\MPSANT}{%
\rm Proc. XXII Int. Symp. on Multiparticle Dynamics, Santiago de Compostela,
Spain, 1992, ed.~A.~Pajares  (\WSCP, 1993)}

\newcommand{\RINGBERG}{%
\rm Proc. Ringberg Workshop on Multiparticle Production, Ringberg Castle,
Germany 1991, eds.~R.C.~Hwa, W.~Ochs and N.~Schmitz  (\WSCP, 1992)\ }
%
\newcommand{\MARBURG}{%
\rm Proc. Int. Workshop on Correlations and Multiparticle Production, Marburg,
eds.~M.~Pl\"umer,  S.~Raha and R.M.~Weiner   (\WSCP, 1991)}
\newcommand{\SANTAFE}{%
\rm Proc. Santa F\'e Workshop Intermittency in High Energy Collisions, 1990,
eds.~F.~Cooper, R.C.~Hwa and I.~Sarcevic (\WSCP, 1991)}

%
%
\def\mode{in comparison with the predictions of  two \MC models.} 
\def\lides{The total error is shown along with the statistical error 
(inner error bars) for each point.
} 
\def\vliet{The error bars show the total uncertainties.}


\renewcommand{\Huge}{\huge}

\title{\vs{4.cm}
\bf Genuine Correlations of Like-Sign Particles \\ 
in   Hadronic {\z} Decays\\
}

\author{\Large The OPAL Collaboration}
\maketitle

\thispagestyle{empty}


\vs{-9.4cm}
\begin{center}
{\Large EUROPEAN LABORATORY FOR NUCLEAR RESEARCH}
\end{center}
\vs{.5cm}
\begin{flushright}     
{\large CERN-EP-2001-070
\\
25th September 2001
}
\end{flushright}

\vs{5.cm} 
     

\begin{abstract}
\noindent

\noindent Correlations among hadrons with the same
electric charge produced in \z\ decays are studied using the high statistics  data
collected from 1991 through 1995 with the \OP detector at LEP.
Normalized factorial cumulants up to fourth order are used to measure genuine particle correlations 
as a function of the size of phase space  domains
 in rapidity, azimuthal angle and transverse momentum.
Both  all-charge  and like-sign particle combinations show
strong positive genuine correlations.
One-dimensional  cumulants initially 
increase  rapidly  with decreasing size of the phase space cells
but saturate quickly.
In contrast,  cumulants in two- and three-dimensional domains 
 continue to increase. 
The strong rise  of the cumulants for all-charge multiplets is increasingly 
driven by that of  like-sign multiplets. 
This points to the likely influence of Bose-Einstein correlations.
Some of the recently proposed algorithms to simulate Bose-Einstein effects,
 implemented in the Monte Carlo  model \PYTHIA, are found to
reproduce reasonably well the measured second- and higher-order
correlations between   particles with the same charge 
as well as those in all-charge particle multiplets.

\end{abstract}
\vs{1.2cm}
\centerline{\Large {\tt }}
\vs{1.cm}

\begin{center}
\em \Large To be submitted to Phys. Lett. B
\end{center}
\nwp
\bigskip
\begin{center}{
G.\thinspace Abbiendi$^{  2}$,
C.\thinspace Ainsley$^{  5}$,
P.F.\thinspace {\AA}kesson$^{  3}$,
G.\thinspace Alexander$^{ 22}$,
J.\thinspace Allison$^{ 16}$,
G.\thinspace Anagnostou$^{  1}$,
K.J.\thinspace Anderson$^{  9}$,
S.\thinspace Arcelli$^{ 17}$,
S.\thinspace Asai$^{ 23}$,
D.\thinspace Axen$^{ 27}$
% ...
}\end{center}\bigskip
\bigskip
$^{  1}$School of Physics and Astronomy, University of Birmingham,
Birmingham B15 2TT, UK
\newline
$^{  2}$Dipartimento di Fisica dell' Universit\`a di Bologna and INFN,
I-40126 Bologna, Italy
\newline
$^{  3}$Physikalisches Institut, Universit\"at Bonn,
D-53115 Bonn, Germany
\newline
% ...
$^{  5}$Cavendish Laboratory, Cambridge CB3 0HE, UK
\newline
% ...
$^{  9}$Enrico Fermi Institute and Department of Physics,
University of Chicago, Chicago IL 60637, USA
\newline
% ...
$^{ 16}$Department of Physics, Schuster Laboratory, The University,
Manchester M13 9PL, UK
\newline
$^{ 17}$Department of Physics, University of Maryland,
College Park, MD 20742, USA
\newline
% ...
$^{ 22}$Department of Physics and Astronomy, Tel Aviv University,
Tel Aviv 69978, Israel
\newline
$^{ 23}$International Centre for Elementary Particle Physics and
Department of Physics, University of Tokyo, Tokyo 113-0033, and
Kobe University, Kobe 657-8501, Japan
\newline
% ...
$^{ 27}$University of British Columbia, Department of Physics,
Vancouver BC V6T 1Z1, Canada
\newline
% ...
Tel Aviv 69978, Israel.


\nwp
\thispagestyle{empty}
 

\nwp

\pagestyle{plain}

\section{Introduction\la{intro}}
Correlations in momentum space 
between hadrons produced in high energy interactions have been
extensively studied over many decades in different contexts~\cite{edw:review}.
Being a measure of event-to-event fluctuations of the number of hadrons in a
phase space domain of size $\Delta$, correlations provide detailed information 
on the hadronisation dynamics, complementary to that derived from  
inclusive single-particle distributions and global event-shape characteristics.

The suggestion in~\cite{bialas1} that  multiparticle  dynamics might possess (multi-)fractal
properties or be ``intermittent'',  emphasized the importance of studying
correlations as a function of the size of domains in momentum space.
A key ingredient for such studies is the normalized
factorial moment and factorial cumulant technique (see Sect.~2), 
which allows statistically meaningful results to be obtained
even for very small phase space cells.

Unlike factorial moments, cumulants of  order $q$ are a direct measure of the stochastic
interdependence among groups  of exactly $q$ particles emitted 
in the same phase space cell~\cite{kendall,Mue71,cumulants}. 
Therefore, they are well suited for the study  of true or ``genuine''
correlations 
between  hadrons.

Whereas earlier work  dealt mainly with
correlations between pairs  {\em {\ie} second-order\/} correlations, 
the use of  factorial cumulants in  high-statistics experiments
has established the presence of genuine correlations
among groups comprising  three or more hadrons, hereafter referred to as  
{\em higher-order\/} correlations.

Experimental results on  hadron correlations are reviewed
in~\cite{edw:review,plo:review}. Except for heavy ion collisions,
where correlations beyond second order are found to be small, as can be
understood from a superposition of independent particle-production 
sources~\cite{sources}, 
in all other types of reactions, significant positive higher-order
correlations are seen.
In two- or three-dimensional (typically rapidity, azimuthal angle,
transverse momentum) 
phase space cells,  they rapidly increase  as the cell-size, $\Delta$,  becomes smaller.
Further studies of the  dependence on particle charge confirm that, as 
conjectured in~\cite{Gyul90},
correlations between hadrons with the same charge play an increasingly
important role
as $\Delta$ decreases, thus  pointing to the influence of  Bose-Einstein (BE) 
interference effects~\cite{hbt,gglp:kopylov}; for a recent review
see~\cite{review:weiner}. 
In contrast,  correlations in multiplets composed of particles with
different charges,
which are more sensitive to
multiparticle resonance decays than like-sign ones, 
tend to saturate in small phase space
domains~\cite{interm:and:bec,lep:higherorderbec}.

Two-particle Bose-Einstein correlations (BEC) have been 
observed in a wide range of multihadronic
processes~\cite{review:weiner}. 
Such correlations were  extensively studied at
LEP~\cite{opal:bec,delphi:rhoshift,bec:asym}. 
Evidence for BEC among groups of more than two identical 
particles has also been 
reported~\cite{lep:higherorderbec,otherthanlep:higherorderbec}.
The subject has acquired  particular  importance  in connection with 
high-precision  measurements of the $W$-boson  mass 
at LEP-II~\cite{sjo,kh}.
For these,  better knowledge of correlations in general is needed,  
as well as realistic Monte Carlo modelling  of BEC~\cite{t1,kittel:review}. 

The OPAL   collaboration recently reported an analysis of 
the domain-size dependence of factorial cumulants  in hadronic Z$^0$ decays, 
using much larger statistics than in any previous experiment~\cite{Oic}.  
In that study, no distinction was made between multiplets composed  of
like-charge particles and those of mixed
charge.
Clear evidence was  seen  for large positive genuine 
correlations  up to fifth order. Hard jet production 
was found to contribute significantly to the observed particle fluctuation patterns.
However, 
Monte Carlo models based on parton showers and string or
cluster fragmentation, gave only a qualitative description of 
the  $\Delta$-dependence of the cumulants. Quantitatively, the model studied,  
which did not explicitly include BE-type correlation effects, underestimated  significantly
correlations between hadrons produced very close together in momentum space.

In the present paper, the high statistics OPAL data collected at and near the Z$^0$ centre-of-mass 
energy are used to  measure cumulants for   multiplets 
of particles with the same charge,  hereafter referred to as ``like-sign cumulants''.
They are compared to  ``all-charge'' cumulants, corresponding to  multiplets comprising particles of any
(positive or negative) charge.
Using the factorial cumulant technique,  results of much higher precision than previously available
are obtained on like-sign correlations up to fourth order.
The role of Bose-Einstein-type effects is studied, using recently
proposed BEC algorithms\footnote{A variety of methods 
which try to simulate BE-effects in Monte Carlo generators have  been developed.  
A detailed exposition of the basic physics issues involved 
 and a review of the presently most popular models  and algorithms 
can be found in~\cite{leif:sjo}.} 
in  the  Monte Carlo event generator \PYTHIA\  for $e^+e^-$
annihilation~\cite{js74}. Proceeding beyond  the usual analyses of two-particle correlations,
we show that, at least 
within the framework of this model, a good  description can be 
achieved of the factorial cumulants up to fourth order
in one-, two- and three-dimensional phase space domains.

The paper is organized as follows. 
Section~\ref{fac:method} explains  the normalized
factorial cumulant  method which allows  
 genuine particle correlations to be measured. 
The OPAL detector, the event selection and the  track selection 
criteria  are detailed in Sect.~\ref{data}. 
The data on factorial cumulants are presented in Sect.~\ref{results}.
They are compared with \PYTHIA\ predictions and, in
particular, with several variants of a model to  simulate the Bose-Einstein effect. 
The results  are summarized in Sect.~\ref{conclusions}.
%
A brief overview of some of the BEC algorithms implemented in \PYTHIA\ 
is given in the appendix.

\section{Factorial cumulant method \label{fac:method}}
To measure genuine multiparticle correlations in  multi-dimensional
phase space cells, 
we use the technique of normalized factorial cumulant moments, $K_q$, or
``cumulants'' for brevity, as proposed in~\cite{cumulants}. 

In the present paper, the cumulants are computed as in a previous OPAL analysis~\cite{Oic}. 
A $D$-dimensional
region of phase space  (defined in Sect.~\ref{data})
is partitioned into $M^D$ cells of equal size $\Delta$.
From the  number of particles counted in each cell, $n_m$
($m=1,\dots,M^D$), event-averaged unnormalized
factorial moments, $\aver{n_m^{[q]}}$, and  unnormalized cumulants,
$k_q^{(m)}$, are
derived, 
using  the relations given {\eg}
in~\cite{kendall}. 
For $q=2,3,4$,  one has
\begin{eqnarray}
k_2^{(m)}&=&\al n_m^{[2]}\ar - \,\al n_m\ar^2,\\
k_3^{(m)}&=&\al n_m^{[3]}\ar - 3\,\al n_m^{[2]}\ar
\al n_m\ar\,
+ 2\, \al n_m\ar ^3\\
k_4^{(m)}&=&\al n_m^{[4]}\ar 
- 4\,\al n_m^{[3]}\ar\, \al n_m\ar
- 3\,\al n_m^{[2]}\ar^2
+12\,\al n_m^{[2]}\ar\,\al n_m\ar^2
-6\, \al n_m\ar^4.
\la{mm}
\end{eqnarray}
Here, $\aver{n^{[q]}}=\aver{n(n-1)\ldots(n-q+1)}$ and 
the brackets $\aver{\cdot}$
indicate that the average over all  events is taken.

Normalized cumulants are calculated using the expression
\begin{equation}
K_q =
({\cal N})^q
\bar{k}_q^{(m)}/
\overline{N_m^{[q]}}.
\la{kmh}
\end{equation}
As proposed  in~\cite{kadija:seyboth}, this form 
is used to correct for statistical bias and non-uniformity  of
the single-particle spectra. 
Here, $N_m$ is the number of particles in the $m$th cell summed over all
$\cal N$ events  in the sample,
$N_m= \sum_{j=1}^{\cal N}(n_m)_j$.
The horizontal bar indicates averaging over the $M^D$ cells in each event,
$(1/M^{D})\sum_{m=1}^{M^{D}}$.

The factorial moment of order $q$ 
of the multiplicity distribution of particles {\em of the same species\/}  
(\eg\ all charged, negatives only,
$\ldots$)
in a  phase space domain   $\Delta$ is equal to the
integral of the $q$-particle inclusive density, $\rho_q$, over that
domain~\cite{edw:review}. 
As in the cluster expansion in statistical mechanics, 
$\rho_q$  
can be decomposed into the sum of contributions from ``accidental''
coincidences of particles in $\Delta$ and the 
true or ``genuine'' correlations.  The latter are denoted here by the
unnormalised cumulants, $k_q$, being the bin-averaged
factorial cumulant functions, or ``correlation functions'', for
short~\cite{Mue71}. 

Whereas $\aver{n^{[q]}}$ depends on all correlation functions of order $1\leq p\leq q$,
$k_q$ is a direct measure of stochastic dependence in multiplets of exactly $q$ particles.
By construction, 
$k_q$  vanishes whenever a particle within the
$q$-tuple is statistically independent of one of the others.
For  Poissonian multiplicity  fluctuations, 
the cumulants of all orders $q>1$ vanish identically. Non-zero cumulants therefore signal the presence of
correlations.

In the following, data are presented for  ``all-charge'' and for ``like-sign'' multiplets.
For the former, the cell-counts $n_m$  are determined using  
all charged particles in an event, irrespective of their charge. 
For the latter,  the number of  positive particles  and the number of  
negative particles in a cell
are counted  separately. The corresponding cumulants  are then averaged to
obtain those
for like-sign multiplets. It is to be noted that the cell-counting
technique does not  allow correlations to be
 measured directly  among  groups composed of particles of different  charge. 
For these, other methods, such  as correlation integrals 
(see {\eg} Sect.~4.8 in~\cite{edw:review}) have to be used.
Nevertheless, cumulants for unlike-sign pairs can be indirectly derived using the equation
\begin{equation}
K_2^{\text{\rm all}}=\frac{1}{2}\, K_2^{\text{\rm ls}}+ \frac{1}{2}\, K_2^{\text{\rm us}},
\label{eq:plusminus}
\end{equation}
which relates second-order cumulants for all-charge pairs to those of
like-sign (ls) and unlike-sign (us) pairs.

\section{Experimental details\la{data}}
The present analysis uses a sample of approximately
$4.1$$\times$$10^6$ hadronic {\z} decays collected from 1991 through 1995.
About 91\% of this sample was taken  at the {\z}; the
remaining part has a centre-of-mass energy, $\sqrt{s}$, 
within $\pm 3$ GeV of the {\z} peak. 

The OPAL detector has been described in detail in~\cite{bib-opal}.
The results presented here are mainly based on the information from the central 
tracking chambers, which consist of a silicon microvertex detector,
a vertex chamber, a jet chamber with 24 sectors each containing 159  axial anode wires,
and outer $z$-chambers to improve the $z$ 
coordinate resolution\footnote{
The right-handed OPAL coordinate system is defined
 with the $z$ axis pointing in the direction of the e$^-$ beam and
 the $x$ axis pointing towards the centre of the LEP ring. 
$r$ is the coordinate normal to the beam axis, 
$\varphi$  the azimuthal angle with respect to the $x$ axis, 
 $\theta$  the polar angle with respect to the $z$-axis.}.
These detectors are located in a 0.435~T axial magnetic field and
measure $p_t$, the track momentum transverse to the beam axis, 
with a precision of
\mbox{$(\sigma_{p_{t}} / p_{t}) = \sqrt{ (0.02)^2 +
(0.0015 \, p_{t})^2 }$~($p_{t}$
 in GeV/$c$)}
for \mbox{$|\cos \theta| < 0.73$}.
%
The jet chamber also measures the specific energy loss of charged particles, 
$\mathrm{dE/dx}$, with a resolution
$\sigma(\mathrm{dE}/\mathrm{dx})/(\mathrm{dE}/\mathrm{dx})\,
\simeq\,3.5\%$
for a track having 159 hits in the jet chamber.
The energy loss  is used to identify charged 
particles~\cite{bib-partid}. 

A sample of over $2$~million  events was
generated with {\sc \JT\!7.4/\PYTHIA6.1}~\cite{js74},
including a full simulation of the detector \cite{bib-opalsim}.
The model parameters were previously tuned to 
OPAL data~\cite{bib-jetset,bib-evshape} but  Bose-Einstein effects were not explicitly
 incorporated.
These events were used to determine 
the efficiencies of track and event selection and for correction purposes.
In addition, for the evaluation  of systematic errors,
 over $1.1$~million events were simulated with \PYTHIA\ including BEC with the
algorithm\footnote{We used 
algorithm BE$_{32}$ (see~\cite{leif:sjo} and the appendix)  
in subroutine {\tt PYBOEI} with parameters 
{\tt MSTJ(51)}=2, 
{\tt MSTJ(52)}=9, 
{\tt PARJ(92)}=1.0,
{\tt PARJ(93)}=0.5~GeV.
 {\tt PARJ(92)} is equal to $\lambda$ in 
Eqs.~(\ref{eq:f2:bec}) and (\ref{eq:f2:bec2}) of the appendix;
the parameter $R$ is given by ${\hbar c}/{\mbox{\tt PARJ(93)}}$.}   
BE$_{32}$.

The event selection criteria are based on the multihadronic event
selection algorithms described in~\ct{Oic}.
It was required that tracks have at least 20 hits in the jet chamber, a first measured point
at a maximum radial distance from the interaction point of 70 cm, 
a minimum transverse momentum  with respect to the beam direction, $p_{t}$, of 0.15~GeV/$c$,
a measured momentum $p$ smaller than  10~GeV/$c$,
a measured polar angle satisfying $|\cos\theta|<0.93$,
and a measured distance of closest approach to the origin 
of less than 5~cm in the $r-\varphi$ plane, and less than 40~cm in the $z$ direction.

The mean energy loss, $\mathrm{dE}/\mathrm{dx}$, of a  track 
was required to be less than 9 keV/cm, thereby rejecting electrons and positrons.
About 97{\%}  of photon conversion pairs 
were rejected, reducing the conversion background 
in the sample to less than 0.1{\%}  of the number of tracks~\cite{bib-partid}.
The fraction of pions in the track sample was estimated from Monte Carlo simulation
to be about 80\%.

Selected multihadron events were required to have at least 5 good tracks,
a momentum imbalance (the  magnitude of the vector sum of the momenta of all charged particles)
 of less than $0.4~\sqrt{s}$ and
the sum of the energies of all tracks (assumed to be pions) greater than 0.2~$\sqrt{s}$.
These requirements provide rejection of background from non-hadronic
{\z} decays, two-photon events, beam-wall and beam-gas interactions. 
In addition, the polar angle of the event sphericity axis,
calculated using tracks that passed the above cuts, as well as 
unassociated electromagnetic and hadronic calorimeter clusters,
had to satisfy \mbox{$\mathrm{|\cos \theta_{\mbox{\small sph}}|} < 0.7$} 
in order to accept only events well contained in the detector.  
A total of about $2.3$$\times$$10^6$ events were finally selected for further analysis. 

The cumulant analysis is performed in the kinematic variables
rapidity, $y$,  azimuthal angle, $\phi$, and the transverse momentum variable,
$\ln{p_T}$, all calculated with respect to the sphericity axis. \begin{itemize}
 \item Rapidity is defined  as $y=0.5\ln [(E+p_{\|})/(E-p_{\|})]$, with $E$ and
$p_{\|}$  the energy (assuming the pion mass) and longitudinal
momentum of the particle, respectively. 
Only particles within the central rapidity region $-2.0\leq y\leq 2.0$ were retained.

 \item In transverse momentum  subspace, the  logarithm of $p_T$, instead of $p_T$ itself, 
is used to eliminate as much as possible the 
strong dependence of the cumulants on cell-size arising from the nearly exponential
shape of the $p_T^2$-distribution. Only particles within the range $-2.4\leq\ln(p_T)\leq0.7$ 
($p_T$ in GeV/{\em c}) were used.

 \item The azimuthal angle, $\phi$, is calculated with respect to the eigenvector
of the momentum tensor having the smallest eigenvalue in the plane
perpendicular to the sphericity axis. The angle $\phi$ spans the interval $0\leq\phi<2\pi$. 
 \end{itemize}

The phase space is partitioned into $M$  bins of equal size for each of the three variables.
$M$ varies between $1$ (full interval) and 400, 
corresponding to a  smallest bin size in one dimension of 
 $\delta y_{\rm min}=0.01$, 
$\delta \phi_{\rm min}=0.9^\circ$
 and $\delta (\ln \pT)_{\rm min}=0.008$, each
significantly larger than  the  experimental resolution of the OPAL detector.

The   cumulants for like-sign and all-charge multiplets, 
measured as a function of $M$, 
have been corrected for  geometrical acceptance,
kinematic cuts, initial-state radiation, 
resolution, secondary interactions and  decays in the detector,
using   correction factors, $U_q(M)$, calculated for all-charge particle combinations 
and evaluated as  in~\ct{Oic} using the {\sc \JT\!\!/\PYTHIA\/} Monte
Carlo without BEC.

As systematic uncertainties, we include the following contributions:
\begin{itemize}
\item 
 The statistical error of the correction factors  $U_q(M)$.
Statistical errors due to the finite statistics of the Monte Carlo
samples are comparable to
those of the data.

\item Track and event selection criteria variations as in  \ct{Oic}. To
this end,
the cumulants have been computed 
changing in turn the following selection criteria: the first
measured point was required to be closer than 40 cm to the beam, the
requirement of the transverse momentum with respect to the beam axis was
removed, the momentum was required to be less than 40 GeV/$c$, the 
track polar angle acceptance was changed to $|\cos \theta|<0.7$,
and the requirement on the mean energy loss was removed. These changes 
modify the results by no more than a few percent in the smallest cells,
 and do not affect the  conclusions.  

\item The difference between  cumulants corrected with 
the factors $U_q(M)$ derived from  Monte-Carlo calculations with and without Bose-Einstein 
simulation. The correction
factors in these two cases differ by at most 5\% in the smallest bins.

\item The difference between  cumulants corrected with 
      $U_q(M)$-factors, calculated for all-charge particle combinations and those calculated for
 like-sign combinations.  The correction factors  coincide within 1\%.
\end{itemize}

The total errors have been calculated by adding the systematic and
statistical \err in quadrature.
It was further verified that our conclusions remain unchanged when events
taken at energies off the {\z} peak are excluded from  the analysis.

\section{Results\label{results}}
\subsection{Like-sign and all-charge cumulants}
The fully corrected normalized cumulants $K_q$ ($q=2,3,4$) 
for all-charge\footnote{The data points for all-charge particle combinations
are the same as those  published in~\cite{Oic} 
but include different systematic uncertainties.}  and 
like-sign  particle multiplets, 
calculated  in one-dimensional ($y$ and $\phi$) (1D),  two-dimensional
$y\times\phi$ (2D) and three-dimensional $y\times\phi\times\ln{p_T}$ (3D) 
phase space cells, are displayed in Fig.~\ref{fig:1} and Fig.~\ref{fig:2}.

From Fig.~\ref{fig:1} it is seen  that, even in 1D, positive 
genuine correlations among groups of two, three and four particles are present: $K_q>0$.
They are substantially stronger in rapidity than in azimuthal angle.
The cumulants increase  rapidly with increasing $M$, the inverse of the bin size,
 for relatively large domains but saturate rather
quickly.
For $K_2$ this behaviour follows from the shape of the second-order 
correlation function
which is known to be 
approximately Gaussian~\cite{edw:review} in the two-particle rapidity
difference
$\Delta\equiv\delta y$.
The rapid rise and subsequent saturation can be understood  from hard gluon jet emission.
For small $M$, the cumulants are sensitive to the large-scale structure of the event, where any
jet non-collinear with the event axis produces a strong fluctuation in the particle density.
With $M$ increasing, the structure inside a single jet is progressively probed.

In contrast to 1D cumulants, those  in 2D and 3D (Fig.~\ref{fig:2}) continue to increase 
towards small phase space cells. Moreover, the 2D and 3D cumulants are of similar
magnitude  at fixed $M$, 
indicating that  the contribution from correlations in transverse momentum  
is small.
This can be understood from  the importance 
of multi-jet production in $e^+e^-$  annihilation, which is most prominently observed 
 in  $ y\times\phi$ space~\cite{Oic}. Indeed, the 1D cumulants in
$p_T$ are found to be close to zero and therefore not shown.

It is to be noted that  oscillations of the cumulants, clearly  seen in the data as well as 
in the Monte Carlo predictions (see \eg\  $K_q(\Phi)$ in Fig.~\ref{fig:1}) are
not due to statistical fluctuations. They arise  from the jet structure
of the events in  rapidity-azimuthal angle subspace, and from the phase space partitioning
 technique used to calculate the cumulants as a function of $M$.

The 1D cumulants of all-charge and of like-sign
multiplets  (Fig.~\ref{fig:1}) show a similar dependence on $M$. 
The latter, however, are significantly
smaller, implying that, for all $M$, correlations 
among particles of opposite charge are important
in one-dimensional phase space projections. Using Eq.~(\ref{eq:plusminus}),  
it is found {\eg} that the one-dimensional   cumulants for  unlike-charge
pairs are larger by  a
factor of $1.6\!-\!1.7$ than those of like-sign pairs for  $M\gtrsim 5$.
This can be expected in general from local charge conservation and 
in particular from resonance decays.

In 2D and 3D (Fig.~\ref{fig:2}), like-sign cumulants increase faster and
 approach the all-charge ones at large $M$. From  Eq.~(\ref{eq:plusminus}), it can be inferred    
that $K_2$ for  unlike-charge pairs remains essentially constant for 
$M$ larger than about 6. Consequently,
as the  cell-size becomes smaller, the rise of all-charge correlations 
is increasingly driven by that  of  like-sign multiplets.
 
Similar features, but based on measurements of {\em factorial moments\/},
 were observed  in other  experiments~\cite{interm:and:bec} and, mainly on qualitative grounds, 
 taken as evidence for the large influence of the  Bose-Einstein 
effect on all-charge correlations 
in small phase space domains\footnote{For a recent critical review we refer to~\cite{interm:bec}.}.
In the next  section,  we add quantitative support to those earlier
observations, 
showing that algorithms which simulate BEC   indeed allow  a quite successful
description of the measurements.

\subsection{Model comparison\label{sec:models}}
In this section,  we compare the cumulant data with predictions of
the \PYTHIA\  Monte Carlo event generator (version 6.158)  without and with Bose-Einstein effects.
Samples of about $10^6$
multihadronic events were generated at the {\z}  energy.
 The model parameters, not related to BEC,  were set at values obtained from a previous tune to OPAL data
on event-shape  and single-particle inclusive   
distributions~\cite{bib-jetset}. In this  tuning,  BE-effects were not
included.

To assess the importance  of BE-type short-range correlations between identical 
particles, and their influence on all-charge cumulants, 
we concentrate on the algorithm BE$_{32}$,
described in the appendix, using parameter values
$\mbox{\tt PARJ(93)}=0.26$~GeV ($R=0.76$~fm) and $\mbox{\tt
PARJ(92)}\equiv\lambda=1.5$.
These values   were determined by varying independently  
$\mbox{\tt PARJ(93)}$ and $\lambda$ within the
range $0.2$--$0.5$ GeV and $0.5$--$2.2$, respectively, 
in steps of $0.05$ GeV and $0.1$, leaving all other 
model-parameters unchanged, until satisfactory agreement with the 
measured cumulants $K_2$ for  like-sign pairs  
was reached\footnote{%
Non-BEC related model-parameters were set at the following values: 
{\tt PARJ(21)}=0.4 GeV,
{\tt PARJ(42)}=0.52 GeV$^{-2}$,
{\tt PARJ(81)}=0.25 GeV,
{\tt PARJ(82)}=1.9 GeV.}. 
Whereas a detailed multi-dimensional best-fit tuning procedure 
was considered to be outside the scope of the paper, 
we find that calculations 
with   $\mbox{\tt PARJ(93)}$ in the range $0.2-0.3$ GeV, and the
corresponding
$\lambda$ in the range $1.7-1.3$,  still provide
an acceptable description of the second-order like-sign cumulants.

Additional  studies have revealed some
 sensitivity of single-particle spectra and event shapes to the inclusion
of BEC in \PYTHIA.
Nevertheless, it was found  that minor variations 
of the QCD and fragmentation parameters are sufficient 
to restore agreement with these data, while the changes in the predicted cumulants remain
well within the systematic uncertainties of the measurement. These parameters were, therefore,
not changed from their  values listed in 
footnote 6.

The dashed  lines in Figs.~\ref{fig:1} and \ref{fig:2} show   \PYTHIA\
predictions
for {\em like-sign\/} multiplets for the model without BEC.
Model and  data agree for small $M$ (large phase space domains), indicating
that  the multiplicity distribution in those regions is  well modelled.
However, for larger $M$, the  predicted cumulants are too small, the
largest deviations occuring in 2D and 3D.
The model predicts negative values for  $K_4(\phi)$ which are not shown.

The solid curves in  Figs.~\ref{fig:1} and \ref{fig:2} show 
predictions for {\em like-sign\/} multiplets using the  BE$_{32}$ algorithm.
Inclusion of BEC leads to a very significant improvement of the data description.
 Not only two-particle
but also higher order correlations  in 1D rapidity space are well  accounted for. 
In $\phi$-space (Fig.~\ref{fig:1}), $K_3$ and especially (the very small)  $K_4$ 
are less well reproduced. Figure~\ref{fig:2} also shows that  the predicted 2D and 3D 
cumulants  agree well with the data.

For more clarity and later reference,
the 1D, 2D and 3D cumulants 
for particle pairs with the same charge are  repeated in Fig.~\ref{fig:7}.
Since BEC occur only
when two identical mesons are close-by in all  three phase space dimensions, 
projection onto lower-dimensional subspaces, such as rapidity 
and azimuthal angle, leads to considerable weakening of the effect. 
Nevertheless, the high precision of the data in    
Fig.~\ref{fig:7} allows to demonstrate  clear sensitivity to the presence 
or absence of BEC in the model, which was much less evident in earlier
measurements~\cite{edw:review}.

Whereas the BE-algorithm used implements  pair-wise BEC only, 
it is  noteworthy that the procedure 
also induces like-sign higher-order correlations  of approximately correct magnitude.
This seems to indicate that  high-order cumulants are, to a large extent,
determined by the second-order one (see further Sect.~\ref{sec:ochs}). 
It is not clear, however, whether the agreement is accidental or  implies
that  the physics of $n$-boson ($n>2$) BE effects is indeed correctly simulated\footnote{For a
recent theoretical discussion of multi-boson BEC, see~\cite{heinz:multi:boson}.}.

The influence of like-sign  BE-type  correlations 
on the correlations  in all-charge multiplets becomes clear from 
Figs.~\ref{fig:3} and \ref{fig:4} where 
all-charge cumulants are compared with 
\PYTHIA\ predictions without BEC, and  for various BE algorithms. 

Large  discrepancies, already discussed in~\cite{Oic} 
and for factorial moments in~\cite{interm:and:bec},
are seen for the model without BEC.  
Inclusion of BE-effects using the  BE$_{32}$ algorithm  
leads to considerably  better agreement, 
in particular in domains of $y$,  $y\times\phi$ and $y\times\phi\times \ln{p_T}$.
In $\phi$,  disagreements similar to those for  like-sign multiplets  are also seen here.
In addition,  $K_2(\phi)$  may be somewhat  overestimated for large $M$.

To assess the sensitivity of the 
cumulants to variations in the BEC algorithms available in \PYTHIA,
we have further considered  the algorithms BE$_\lambda$ and BE$_0$ (see
appendix).

Using the same parameter values as for BE$_{32}$,
we observe that  BE$_\lambda$ slightly  
underestimates $K_2(y)$ and overestimates  $K_2(\phi)$
for like-sign pairs (Fig.~\ref{fig:7}), 
whereas  the results  coincide with those from BE$_{32}$ in 2D and 3D. 
For all-charge multiplets (Figs.~\ref{fig:3} and \ref{fig:4}),  
the predicted cumulants  generally fall below those 
for  BE$_{32}$, except for  $K_3$ and $K_4$ in 2D and 3D, where the  differences  are small.
We note, in particular, that the  
predicted $K_2(y)$  (Fig.~\ref{fig:3}) essentially coincides with the \PYTHIA\ results
{\em without\/} BEC.
The differences with respect to  BE$_{32}$ are related 
to the different  pair-correlation functions $g_2(Q)$ 
(Eqs.~\ref{eq:f2:bec} and  \ref{eq:f2:bec2} of the appendix) used in the
algorithms.
Although a different choice of the parameters $R$ and $\lambda$ 
may improve the agreement with the data, we have not attempted such fine-tuning.

In $W$-mass studies using  the fully hadronic decay channel of the reaction 
$e^+e^-\to W^+W^-$,  and assuming that
pairs of pions from different $W$'s are fully affected by BEC, algorithm  BE$_{32}$
is found to introduce a negative mass-shift, whereas BE$_{\lambda}$ generates a 
positive shift~\cite{leif:sjo}. The  mass-shift occurs since the BE-induced momentum changes increase
the likelihood that soft particles in an event are assigned to the wrong jet. This, in turn,
depends on the strength of the particle correlations in 3D momentum space.
Since BE$_{32}$ and   BE$_{\lambda}$ provide a
reasonable  description of the domain-size dependence of   all-charge cumulants in 3D, 
both algorithms should be considered in $WW$ studies. 

Finally, we consider the model predictions based on the  algorithm BE$_0$
(dash-dotted curves in the figures) for the same parameter values as quoted above.
For like-sign pairs (Fig.~\ref{fig:7}), $K_2(y)$  and especially $K_2(\phi)$ are overestimated.
This is also the case for  $K_2(\phi)$  for all-charge pairs  shown in Fig.~\ref{fig:3}.
In contrast, all-charge  higher-order cumulants differ little from those
obtained with   BE$_{32}$.
It should be noted that the  BE$_0$  algorithm, contrary to  BE$_{32}$ and BE$_{\lambda}$,
enforces   energy conservation  by a global
rescaling of all final-state hadron momenta. This procedure  affects the full hadronic final state and
induces a large artificial shift in the $W$-mass when applied to the
reaction  $e^+e^-\to W^+W^-\to \mbox{hadrons}$. It is, for that reason, 
disfavoured by the authors of~\cite{leif:sjo}.

Traditional studies of the Bose-Einstein effect most often consider only  the 
second-order correlation function for like-sign particle pairs, which 
is measured as a function of ${Q^2}$, 
the square of the difference in four-momenta
$p_i$ ($i=1,2$) of particles in the pair. The variable $Q^2$ is related to
the 
3D cell-size used in this paper, via the approximate relation~\cite{q:formula}:
\begin{equation}
Q^2\approx m^2\left[
\beta^2 \delta\phi^2 +(1+\beta^2)\delta y^2 +
\frac{\beta^4\delta x^2}{(1+\beta)^2}
\right],
\label{eq:cellsize}
\end{equation}
valid for small domains in $y$, $\phi$ and $\ln{p_T}$. 
Here $\beta=p_T/m$, $m$ is the particle mass and $\delta x=\ln(p_{T1}/p_{T2})$. 
It follows that a small cell-size in 3D 
corresponds to small $Q^2$. The reverse is not true.
Equation~(\ref{eq:cellsize})~implies that  a measurement  of $K_2$ versus $M$ in 3D 
is roughly equivalent to that of  the inclusive two-particle $Q$-distribution. 

As a  check of the consistency of our results, we have  compared predictions for  the inclusive 
two-particle $Q$-distribution of like-sign pairs
(single-particle rapidity restricted to the interval $-2\leq y\leq 2$) 
from the BE$_{32}$ model to that extracted from the data
and  fully corrected for detector effects.
For the values of the BEC parameters $R$ and $\lambda$ quoted above, 
satisfactory agreement is obtained (not shown).

In recent work, several LEP experiments have analyzed 
 the second-order like-sign correlation function in terms of individual
components of the
four-vector  $Q$~\cite{bec:asym}. 
If interpreted as a measurement of the space-time extent of the
particle emitting source, such results show that the emission volume  has an elongated shape 
with respect to the event axis. The BE algorithms discussed here, which treat
all components of $Q$ 
symmetrically,
are unable to reproduce these measurements~\cite{bec:fial}.
It is therefore possible, although to be verified,
that  the measured anisotropy  of the two-particle 
correlation function may be responsible for 
some of the \PYTHIA\ model discrepancies, {\eg} for 1D cumulants in $\phi$ domains,
observed in the present analysis.

In summary, a comparison with   \PYTHIA\ predictions
shows that 
short-range correlations of the BE-type 
are needed, at least in this model,
to reproduce  the magnitude and the $\Delta$-dependence of the cumulants
for like-sign multiplets. This further leads to a much improved description of not only
two-particle but also of higher-order  correlations in all-charge multiplets.
Since Bose-Einstein
correlations are a well-established phenomenon in multiparticle
production, it is likely that
the above conclusion has wider validity than the model from which it was
derived.

\subsection{The Ochs-Wosiek relation for cumulants\label{sec:ochs}}
The success of the \PYTHIA\  model with BEC in predicting both  the magnitude and
domain-size dependence of  cumulants, has led us to consider 
the inter-dependence of these quantities in more detail. 

In Fig.~\ref{fig:5} we plot $K_3$ and $K_4$   in 2D and 3D, 
for each value of $M$ as a function of $K_2$.
We observe that the 2D and 3D data for all-charge, 
as well as for like-sign multiplets
follow approximately,  within errors, the same functional dependence. 
The solid lines is a simple fit to the function 
\begin{equation}
\ln{K_q}=a_q+ r_q\ln{K_2}.\label{eq:ochs:wosiek}
\end{equation}
The fitted slope values are
$r_3=2.3$ and $r_4=3.8$.
This is evidence that
the slope $r_q$ 
increases with
the order of the cumulant. The $q$-dependence of the slopes is of particular interest
in  multiplicative cascade models of hadronisation, and indicative of the mechanism causing 
scale-invariant fluctuations~\cite{bialas1,BrPe91}.

Figure~\ref{fig:5} 
 suggests  that the {\em cumulants\/} of different orders obey  simple so-called ``hierarchical'' 
relations,  analogous to the  Ochs-Wosiek relation,
first established  for {\em factorial moments\/}~\cite{Ochs:Wosiek:relation}.
Interestingly,  all-charge as well as   like-sign multiplets are  seen to follow, within errors,
 the  same functional dependence.
Hierarchical relations of similar type  are commonly encountered, or conjectured, in various branches
of many-body physics (see {\eg}~\cite{cumulants}), but a satisfactory explanation
within particle production phenomenology or QCD remains to be found.

Simple relations among the cumulants of different orders exist for certain
probability distributions, such as the Negative Binomial distribution, which often  
parameterizes successfully
the multiplicity distribution of hadrons in restricted phase space regions~\cite{neg:bin}.
For this distribution, one has $K_q=(q-1)!\,K_2^{q-1}$ ($q=3,4,\dots$),
showing 
that the cumulants are here solely determined by  $K_2$.
This relation is shown in Fig.~\ref{fig:5} (dashed line). Comparing to the data,
we  conclude that the  multiplicity distribution of all charged particles, 
as well as that of like-sign particles, deviates
strongly from a  Negative Binomial in small phase space domains. For further discussion, 
in the present context, 
of this and other much studied  multiplicity distributions we refer to~\cite{edward}.

The Ochs-Wosiek type of relation 
exhibited by the data in Fig.~\ref{fig:5} may  explain
why the BE algorithms in \PYTHIA\   generate higher-order correlations of
(approximately) the correct magnitude. Assuming that the hadronization 
 dynamics is such that  higher-order correlation functions can be constructed from
second-order correlations only,
methods that are designed to ensure agreement with the two-particle correlation function,
could
then automatically generate  higher-order ones of the correct magnitude.

\section{Summary and conclusions\label{conclusions}}
In this paper we have presented a comparative study of 
 like-sign and all-charge genuine correlations
between two and more hadrons produced in $e^+e^-$ annihilation at the $Z^0$ energy.
The high-statistics data on hadronic $\mathrm{Z}^0$ decays  
recorded  with the OPAL detector from 1991 through 1995
were used  to measure normalized  factorial cumulants as a function of the domain size, $\Delta$,  in
$D$-dimensional domains ($D=1,2,3$) in rapidity, azimuthal angle and (the logarithm of) 
transverse momentum, defined  in the event sphericity frame.

Both  all-charge  and like-sign multiplets show
strong positive genuine correlations up to fourth order.
They are stronger in rapidity than in azimuthal angle.
One-dimensional  cumulants initially 
increase  rapidly  with decreasing size of the phase space cells
but saturate rather quickly.
In contrast,  2D and especially 3D cumulants 
 continue to increase and exhibit intermittency-like behaviour. 

Comparing all-charge and like-sign multiplets  in 2D and 3D phase space cells,
we observe that the rise  of the cumulants for all-charge multiplets is increasingly 
driven by that of  like-sign multiplets  as $\Delta$ becomes smaller. 
This points to the likely influence of Bose-Einstein correlations.

The 2D and 3D cumulants $K_3$ and $K_4$, 
considered as a function of $K_2$, 
 follow  approximately a linear relation of the Ochs-Wosiek  
type: $\ln K_q\sim \ln K_2$, independent of $D$ and the same
for all-charge and for like-sign particle groups.
This suggests that, for a given  domain $\Delta$,  
correlation functions of different orders are not
independent but determined,   to a large extent, by  two-particle correlations. 

The data have been compared  with predictions from the Monte Carlo event 
generator \PYTHIA, previously tuned to single-particle  and
event-shape OPAL data. 
The model  describes well
dynamical fluctuations in large phase space domains, {\eg} caused by jet production, 
and shorter-range correlations attributable to resonance decays.
However, the results of this paper, 
together with earlier  less precise data, show that these ingredients alone 
are insufficient 
to explain the magnitude and domain-size dependence
 of the factorial cumulants. 
To achieve a  more satisfactory data description, 
short-range correlations of the Bose-Einstein type 
between  identical particles need to be included.

The importance of BE-type effects has been studied 
using  algorithms implemented in \PYTHIA.
We find that the model BE$_{32}$ 
is  able to simultaneously account for the
magnitude and $\Delta$-dependence of  like-sign 
as well as of all-charge cumulants. Other models, BE$_0$ and BE$_\lambda$,
when using  the same parameters as for BE$_{32}$, provide also a reasonable
description of the data.
Although the algorithms implement pair-wise BEC only,
 surprisingly good  agreement with the measured  third-  and fourth-order
cumulants is observed. This  could be a consequence of the Ochs-Wosiek type of inter-relationship
between cumulants of different orders, exhibited by the data.
\medskip
\bigskip\bigskip\bigskip
\appendix
\par
Acknowledgements:
\par
We particularly wish to thank the SL Division for the efficient operation
of the LEP accelerator at all energies
 and for their close cooperation with
our experimental group.  We thank our colleagues from CEA, DAPNIA/SPP,
CE-Saclay for their efforts over the years on the time-of-flight and trigger
systems which we continue to use.  In addition to the support staff at our own
institutions we are pleased to acknowledge the  \\
Department of Energy, USA, \\
National Science Foundation, USA, \\
Particle Physics and Astronomy Research Council, UK, \\
Natural Sciences and Engineering Research Council, Canada, \\
Israel Science Foundation, administered by the Israel
Academy of Science and Humanities, \\
Minerva Gesellschaft, \\
Benoziyo Center for High Energy Physics,\\
Japanese Ministry of Education, Science and Culture (the
Monbusho) and a grant under the Monbusho International
Science Research Program,\\
Japanese Society for the Promotion of Science (JSPS),\\
German Israeli Bi-national Science Foundation (GIF), \\
Bundesministerium f\"ur Bildung und Forschung, Germany, \\
National Research Council of Canada, \\
Research Corporation, USA,\\
Hungarian Foundation for Scientific Research, OTKA T-029328, 
T023793 and OTKA F-023259.\\

\newpage
\section*{Appendix: Modelling BEC in PYTHIA}\label{pythia:bec:model}
As emphasized  in~\cite{leif:sjo}, no rigorous method exists to date   
which would
allow to account for the  Bose-Einstein symmetrization of the production amplitudes
in Monte Carlo event generators.
The algorithms implemented in the Monte Carlo event generator  
\PYTHIA~\cite{js74}  
which simulate the BE effect, are all based on introducing BEC
as local shifts of final-state particle momenta among pairs of 
identical particles. They differ only in the way global energy and momentum conservation is ensured.
Other, so-called ``global-weight'' methods, have 
been proposed wherein a single weight-factor is assigned
to each event, calculated on the basis of a model for the identical-pair two-particle
 correlation function~\cite{bo:ringner,global:weight:models}.

The \PYTHIA\  algorithms take the hadrons produced by the string
fragmentation, 
where no BE effects are present, and shift the momenta of mesons $i$ and $j$ such that the inclusive
distribution of the relative separation $Q$ of identical pairs is enhanced by a factor
$g_2(Q)\geq1$. The latter, as used here, is parameterized with the phenomenological 
form\footnote{In naive BE models, 
based on the analogy with optics and assuming a {\em static\/} incoherent emission
source, $R$ is related to the size of the source; $\lambda$  quantifies the
strength of the BE effect.}
\begin{equation}
g_2(Q)=1+ \lambda\,e^{-R^2Q^2},
\label{eq:f2:bec}
\end{equation}
where $Q$  is the difference in four-momenta of the  pair, $Q^2=-(p_1-p_2)^2$.

Short-lived resonances like the $\rho$ and $K^\ast$ are allowed to decay
before the BE procedure is
applied, while decay products of  longer-lived ones 
(width $\Gamma<20$~MeV/$c^2$) are not affected. 
The procedure also influences groups of non-identical particles causing
{\eg\/} a shift
of the $\rho^0$ mass peak~\cite{opal:rhoshift,delphi:rhoshift}.

In  this paper we compare the data to simulations based on the 
\PYTHIA\  algorithms BE$_0$, BE$_{32}$ and BE$_\lambda$, 
following the nomenclature in~\cite{leif:sjo}. 

{\em In the  BE$_0$ algorithm\/}, momentum is conserved exactly, but
energy conservation is
explicitly broken in the treatment of individual particle pairs. It is
restored only by a global
rescaling of all final-state hadron momenta, thus affecting the full
hadronic final state.

The other algorithms avoid global energy rescaling by introducing  additional  momentum shifts
for some pairs of particles, ($k$, $l$),  not necessarily identical, 
in a local region around each identical pair ($i$, $j$). 

{\em The  algorithm $\BE_{32}$\/}, which is applied to identical pairs only, 
 is based on the  ansatz
\begin{equation}
 g_2(Q) = \left\{ 1 + \lambda \exp(-Q^2R^2) \right\}
 \left\{ 1 + \alpha \lambda \exp(-Q^2R^2/9) 
 \left( 1 - \exp(-Q^2R^2/4) \right) \right\}, \label{eq:f2:bec2}
\end{equation}
($\alpha$ is an adjustable parameter) 
which  attempts to mimic effects due to oscillations  below and above unity of the
pair-weights, as found in some models~\cite{bo:ringner}.

{\em In the  algorithm  $\BE_{\lambda}$\/},
the form (\ref{eq:f2:bec}) of $g_2(Q)$ is retained.
For each pair of identical
particles ($i,j$), a pair of non-identical particles, ($k,l$),
neither identical to $i$ or $j$, is found  close to  ($i,j$). 
 For each momentum shift among particles $i$ and $j$, a corresponding
shift among the particles $k$ and $l$ is found so that the total energy and momentum in the
$i,j,k,l$ system is conserved. 
As a measure of  ``closeness'',  $\BE_{\lambda}$ uses the so-called 
$\lambda$-measure~\cite{cascades:fractal}, related to the string length
in the Lund
string fragmentation framework. 

A given particle is likely to belong to several identical pairs in an event. 
The net shift in particle momenta and energies therefore depends on the complete configuration of
all identical particles. This introduces complex effects, not only among identical 
pairs but also  among unlike-sign pairs  and higher multiplicity multiplets of nearby  particles.  
As a result,  genuine higher-order correlations emerge.
For a full description of the BE algorithms mentioned above,  ref.~\cite{leif:sjo} should be consulted.

\newpage



\textheight=25.cm
\nwp

\begin{figure}
\vs{7cm}
\epsfysize=12cm
\epsffile[45 150 200 500]{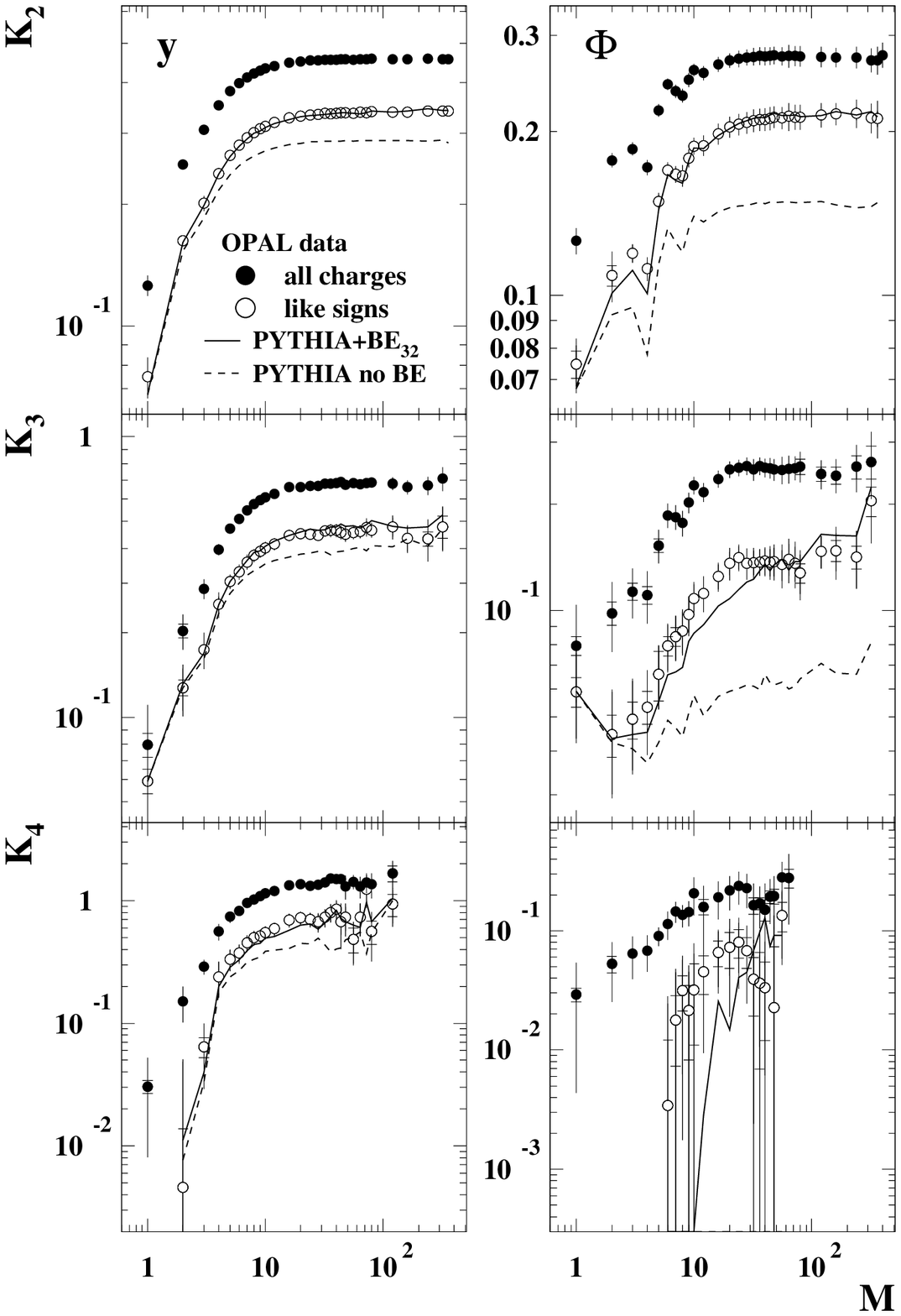} 
\vs{1.cm}
\caption{\it  The cumulants $K_q$  in  one-dimensional domains of
rapidity ($y$) and azimuthal angle ($\Phi$) for all charged hadrons (solid
symbols) 
and for  multiplets of like-sign particles (open
symbols), versus $M$.
 Where two error-bars are shown, inner  ones are  statistical, and 
outer ones are statistical  and 
systematic errors added in quadrature.
The lines  connect   Monte Carlo predictions 
for like sign cumulants from \PYTHIA\ without BEC
(dashed) 
and with BEC (full) simulated with algorithm  
$\mbox{BE}_{32}$  \protect\cite{leif:sjo} (see text).%
} 
\la{fig:1}
\end{figure}

\begin{figure}
\vs{7cm}
\epsfysize=12cm
\epsffile[45 150 200 500]{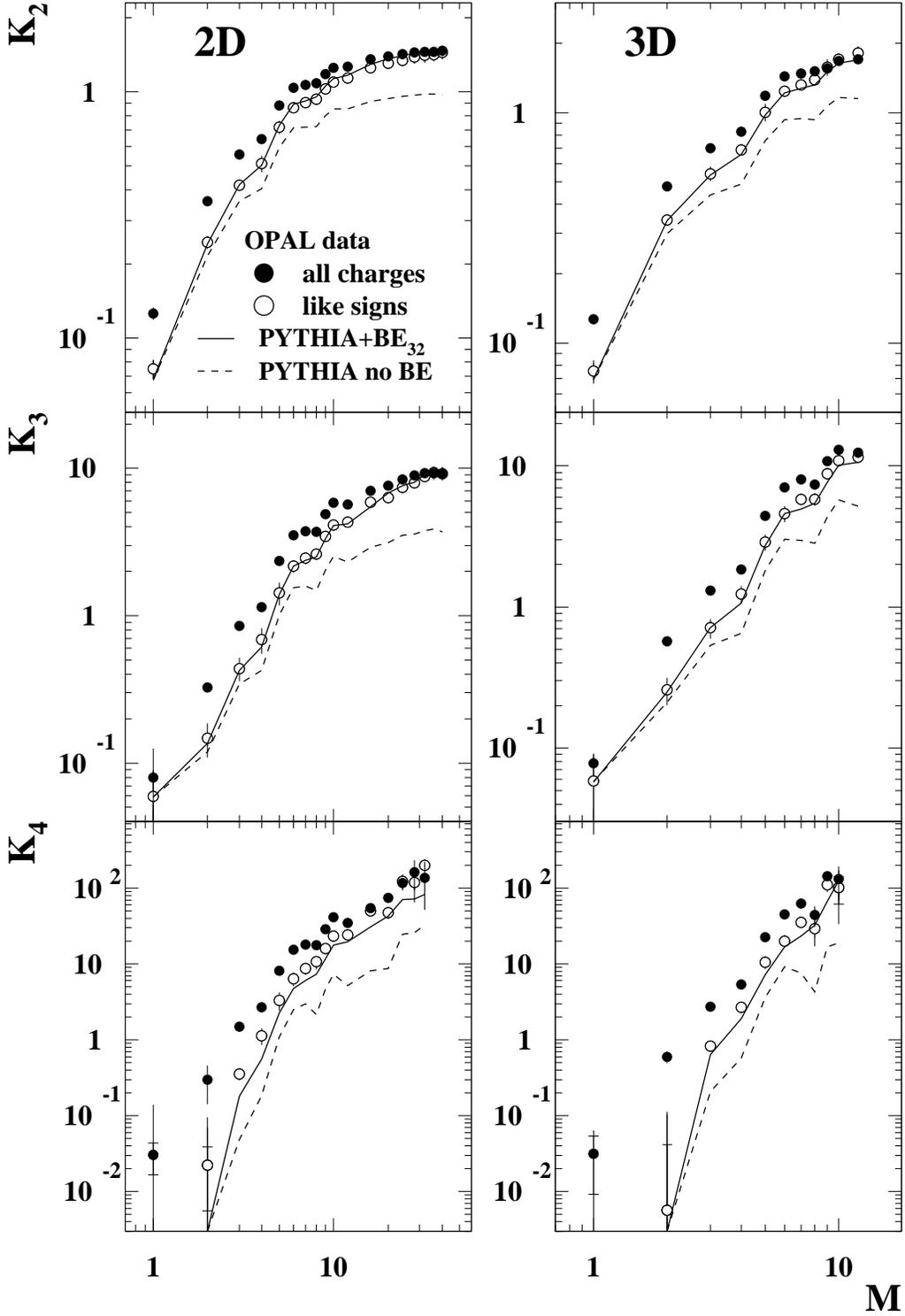} 
\vs{1.cm}
\caption{\it 
 The cumulants $K_q$  in 
 two-dimensional $\Delta y\times\Delta\Phi$ (2D) 
and 
three-dimensional $\Delta y\times\Delta\Phi\times\Delta\ln p_T$  (3D) domains
 for all charged hadrons (solid symbols) 
and for  multiplets of like-sign particles 
(open symbols), versus  $M$. 
 Where two error-bars are shown, inner  ones are  statistical, 
and outer ones are
statistical  and 
systematic errors added in quadrature.
The lines  connect   Monte Carlo predictions from \PYTHIA\ without BEC
(dashed) 
and with BEC (full) simulated with algorithm  
$\mbox{BE}_{32}$ \protect\cite{leif:sjo} (see text).%
 } 
\la{fig:2}
\end{figure}

\begin{figure}
\vs{7cm}
\epsfysize=11cm
\epsffile[45 150 200 500]{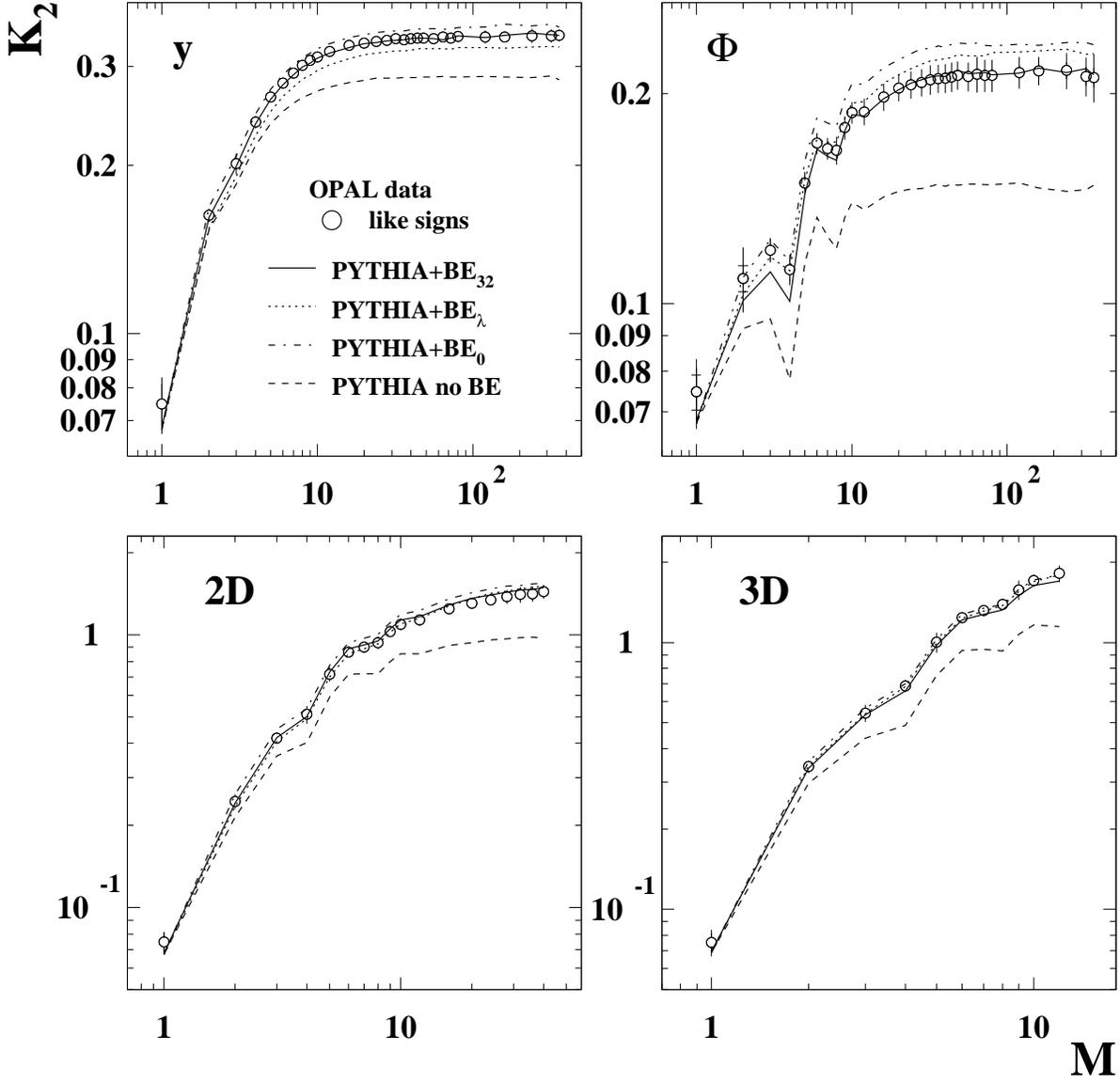} 
\vs{2.cm}
\caption{\it  The cumulants $K_2$  for  like-sign pairs
 in one-dimensional domains of rapidity ($y$) and azimuthal angle ($\Phi$), and in
two-dimensional $\Delta y\times\Delta\Phi$ (2D)
 and three-dimensional  $\Delta y\times\Delta\Phi\times\Delta\ln p_T$  (3D) domains 
versus $M$.
The error-bars show statistical  and systematic errors added in quadrature.
The lines connect  Monte Carlo predictions from \PYTHIA, without BEC
and with various
Bose-Einstein algorithms\protect\cite{leif:sjo} (see text).  
} 
\la{fig:7}
\end{figure}

\begin{figure}
\vs{7cm}
\epsfysize=12cm
\epsffile[45 150 200 500]{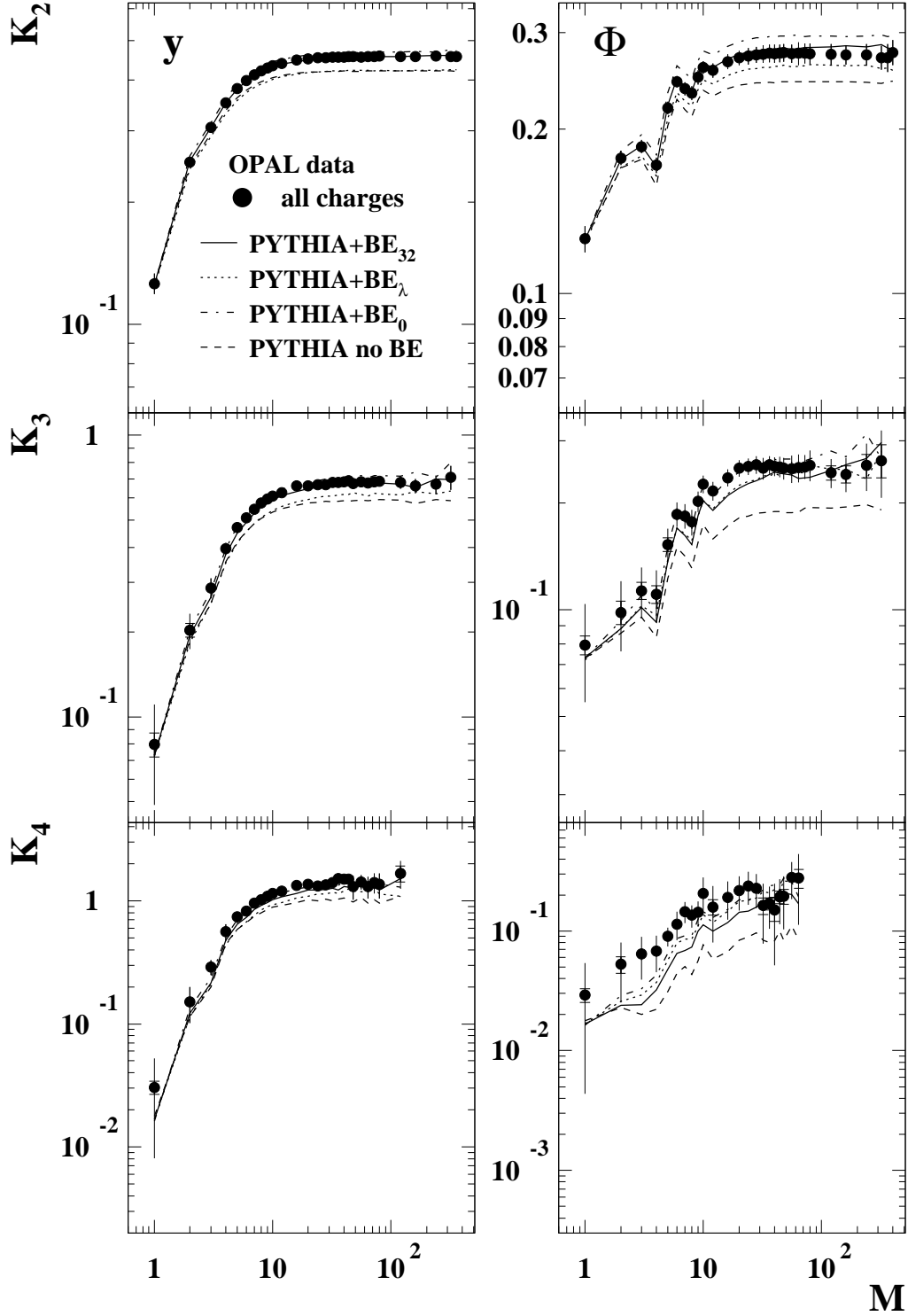} 
\vs{1.cm}
\caption{\it 
 The cumulants $K_q$  in  one-dimensional domains of
rapidity ($y$) and  azimuthal angle ($\Phi$)
for all charged hadrons, as in Fig.~\ref{fig:1}, 
versus $M$.
 Where two error-bars are shown, inner  ones are  statistical, and
outer ones are statistical  and systematic errors added in quadrature.
The lines connect  Monte Carlo predictions from \PYTHIA, without BEC
and with various
Bose-Einstein algorithms\protect\cite{leif:sjo} (see text).  
}
\la{fig:3}
\end{figure}

\begin{figure}
\vs{7cm}
\epsfysize=12cm
\epsffile[45 150 200 500]{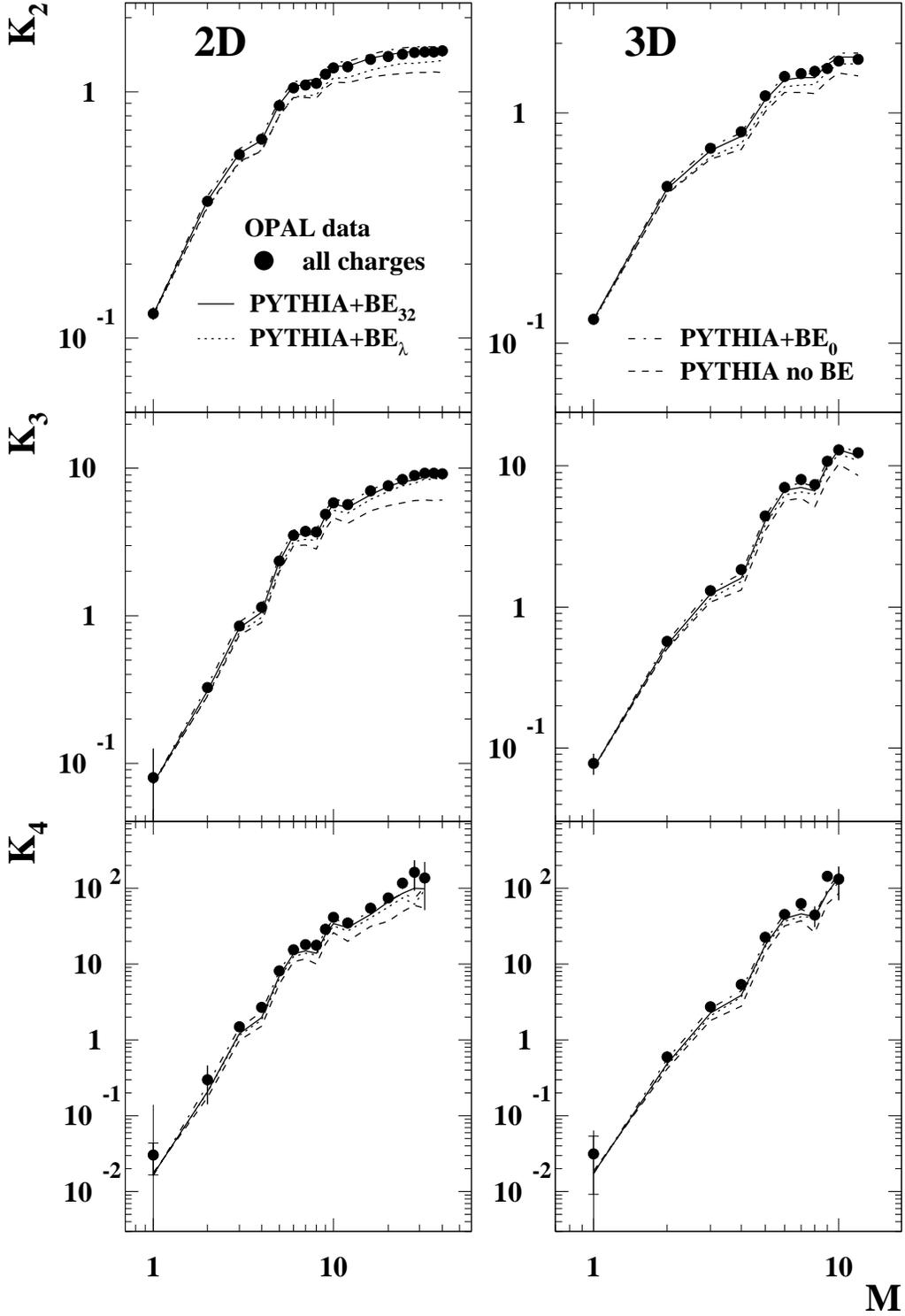} 
\vs{1.cm}
\caption{\it 
 The cumulants $K_q$   
in  two-dimensional $\Delta y\times\Delta\Phi$ (2D) 
and three-dimensional $\Delta y\times\Delta\Phi\times\Delta\ln p_T$  (3D) domains 
for all charged hadrons as in Fig.~{\rm\ref{fig:2}}, versus $M$.
 Where two error-bars are shown, inner  ones are  statistical, 
and outer ones are statistical  and 
systematic errors added in quadrature.
The lines connect  Monte Carlo predictions from \PYTHIA, without BEC
and with various
Bose-Einstein algorithms\protect\cite{leif:sjo} (see text).  
} 
\la{fig:4}
\end{figure}

\begin{figure}
\vs{7cm}
\epsfysize=11cm
\epsffile[45 150 200 500]{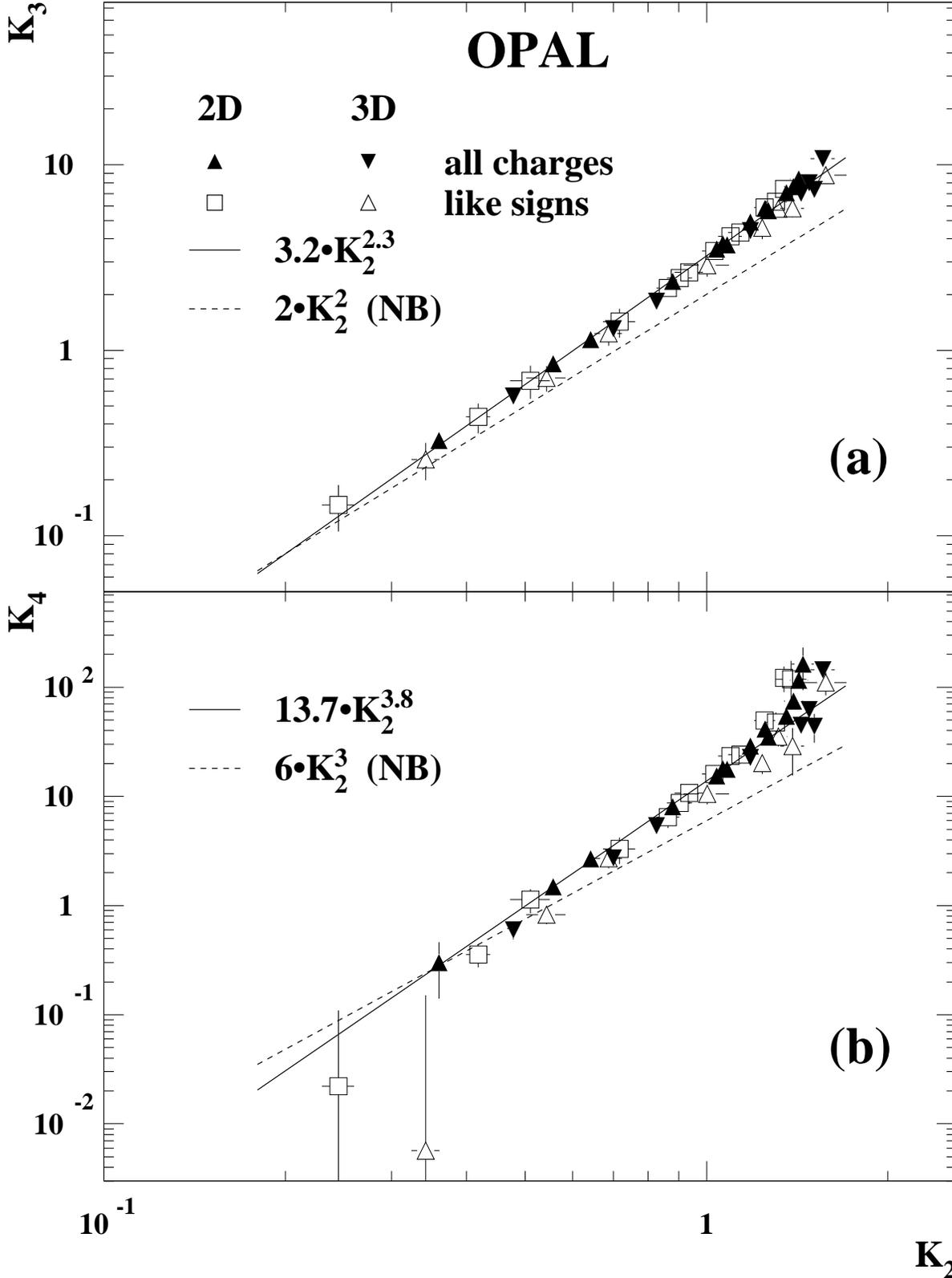} 
\vs{2.cm}
\caption{\it 
The Ochs-Wosiek plot 
in two-dimensional $\Delta y\times\Delta\Phi$ (2D) 
and three-dimensional  $\Delta y\times\Delta\Phi\times\Delta \ln p_T$ (3D) 
domains 
for all charged hadrons (solid symbols) and for multiplets of like-sign
particles  (open symbols). 
The dashed line shows  the function,  $K_q=(q-1)!\,K_2^{q-1}$ ($q=3,4$), 
 valid for
a Negative Binomial multiplicity distribution (NB) in  each phase space cell.
The solid line shows a fit with  Eq.~(\ref{eq:ochs:wosiek}).
} 
\la{fig:5}
\end{figure}

\end{document}